\documentclass[10pt,twocolumn,a4paper]{article}

\usepackage[left=20mm, top=20mm, right=20mm, bottom=20mm, nohead, foot=10mm]{geometry}

\usepackage[utf8]{inputenc}
\usepackage[english]{babel}

\usepackage[affil-it]{authblk}

\usepackage{widetext}

\usepackage{caption}

\usepackage{amsmath,amssymb}

\usepackage{xcolor}
\usepackage[normalem]{ulem}

\definecolor{colorA}{rgb}{0.0, 0.0, 0.5}

\definecolor{colorB}{rgb}{0.5, 0.0, 0.0}

\definecolor{colorC}{rgb}{0.0, 0.5, 0.0}

\definecolor{colorD}{rgb}{0.5, 0.0, 0.5}

\newcommand{\CalA}{\mathcal{A}}
\newcommand{\CalC}{\mathcal{C}}
\newcommand{\CalD}{\mathcal{D}}

\usepackage{titlesec}
\titleformat{\section}[block]{\normalfont\sffamily}{}{.5em}{\bfseries}
\titleformat{\subsection}[runin]{\normalsize\bf}{}{0em}{\bfseries}

\usepackage{hyperref}
\hypersetup{colorlinks=true, linkcolor=[rgb]{0.6 0 0}, citecolor=[rgb]{0 0.4 0}, urlcolor=[rgb]{0 0 0.6}}

\usepackage{graphicx}

\usepackage{csquotes}

\begin{document}

\title{\textit{In-situ} pseudopotentials for electronic structure theory}

\author[1]{\normalsize K.Bj\"ornson}
\author[2]{\normalsize J.M.Wills}
\author[3]{\normalsize M.Alouani}
\author[1]{\normalsize O.Gr\aa n\"as}
\author[1]{\normalsize P.Thunstr\"om}
\author[1]{\normalsize Chin Shen Ong}
\author[1,4]{\normalsize O.Eriksson}

\affil[1]{\small
Department of Physics and Astronomy, Uppsala University, Box 516, SE-75120 Uppsala, Sweden
}

\affil[2]{\small
Los Alamos National Laboratory, Los Alamos, NM 87545, USA 
}

\affil[3]{\small
Universit\'e de  Strasbourg, Strasbourg, Institut de Physique et Chimie des Mat\'eriaux de Strasbourg, UMR 7504  CNRS-UNISTRA, France
}

\affil[4]{\small
School of Science and Technology, \"Orebro University, Fakultetsgatan 1, SE-701 82 \"Orebro, Sweden
}

\date{}

\maketitle

\begin{abstract} 
We present a general method of constructing \textit{in-situ} pseodopotentials from first principles, all-electron, full-potential electronic structure calculations of a solid. The method is applied to bcc Na, at equilibrium volume. The essential steps of the method involve (i) calculating an all-electron Kohn-Sham eigenstate. (ii) Replacing the oscillating part of the wavefunction (inside the muffin-tin spheres) of this state, with a smooth function. (iii) Representing the smooth wavefunction in a Fourier series, and (iv) inverting the Kohn-Sham equation, to extract the pseudopotential that produces the state generated in steps (i)-(iii). It is shown that an \textit{in-situ} pseudopotential can reproduce an all-electron, full-potential eigenvalue up to the sixth significant digit. A comparison of the all-electron theory, \textit{in-situ} pseudopotential theory and the standard nonlocal pseudopotential theory demonstrates good agreement, e.g., in the energy dispersion of the 3$s$ band state of bcc Na.
\end{abstract}  

\section{Introduction}

The electronic structure of solids within the density functional theory (DFT) has
been solved by a variety of methods, such as the  linear combination of atomic
orbitals (LCAO)~\cite{ZPhys.52.555}, the Korringha-Kohn-Rostocker (KKR) Green's
function method~\cite{Physica.13.392, PhysRev.94.1111}, the all-electron linear
muffin-tin orbitals (LMTO) and  the linear augmented plane waves 
(LAPW)~\cite{PhysRevB.12.3060, skriver1984lmto, Wills2010} and the pseudopotential 
method~\cite{Heine1970, Cohen1970, Cohen1988, RevModPhys.64.1046, Chelikowsky2011}.
One key difference between the all-electron and the pseudopotential methods is
in their treatments of core electrons. Whereas in all-electron methods, the
core electrons are explicitly included in the calculations, pseudopotential
methods replace the potentials from the core states in the
one-electron Schr\"odinger (or Kohn-Sham) equation with an effective smooth
potential, known as the pseudopotential. This allows the pseudopotential
methods to replace the valence
states with smooth pseudowavefunctions, which have fewer nodes than the
all-electron wavefunctions but the same eigenenergies.

This approach has its roots in the ideas of Fermi and Hellman more than
eighty years ago~\cite{NuevoCimentoB.11.157, JChemPhys.3.61}, with the rigorous
formulation of the theory for a solid taking place twenty years
after that~\cite{Antoncik1954, JPhysChemSolids.10.314, PhysRev.116.287,
PhysRev.118.1153}. A practical method of predicting energy band structures for
semiconductors was only achieved upon the development of the empirical
pseudopotential method (EPM)~\cite{Cohen1970, Cohen1988}, in which the
pseudopotential is fitted to experimental band structures, establishing
the validity of the energy band concept for solids in general. Nonetheless,
empirical pseudopotentials are not always transferable between systems of
different chemical environments since their suitability for a particular system
depends on the similarity of that environment to the experimental environment
to which the empirical pseudopotential was fitted~\cite{Cohen1988,
Chelikowsky2011}. Consequently, the use of pseudopotentials constructed from
first principles, i.e., \textit{ab initio} pseudopotentials, has become
widespread in modern-day electronic structure research. \textit{Ab initio}
pseudopotentials can be norm-conserving~\cite{Hamann1979,
PhysRevB.26.4199} or ultrasoft~\cite{PhysRevB.41.7892, PhysRevB.41.5414}. In
the projector augmented wave (PAW) approach~\cite{PhysRevB.50.17953,
PhysRevB.59.1758}, pseudopotential operators are also used, but information
regarding the nodal structure of the all-electron wavefunctions in the core
region is retained. Several excellent papers and reviews have already discussed
many necessary details of the \textit{ab initio} pseudopotential approach~\cite{ChemPhysChem.12.3143, RevModPhys.64.1046, PhysRevLett.48.1425}. Here, we will only reiterate some of the pertinent points.

The first step in constructing an \textit{ab initio} pseudopotential usually
involves  solving the Kohn-Sham equations for a free isolated atom to obtain its
all-electron eigenvalues and wavefunctions. In the second step, one constructs
a smooth pseudowavefunction, which has a radial component that is identical to the
radial component of the all-electron wavefunction outside a chosen cutoff
radius, $r_c$, but is smooth and nodeless inside this radius.
Finally, one asks if there exists a potential (i.e., a pseudopotential) that
when used together with the kinetic-energy operator to construct the Kohn-Sham
Hamiltonian, produces this pseudowavefunction as its eigenstate upon
diagonalization, while retaining the same all-electron energy eigenvalue 
of the free atom. This is done by inverting the Schr\"odinger equation.
The pseudopotential approach takes
advantage of the fact that the core electrons do not play an important role in
the formation of chemical bonds between atoms~\cite{JChemPhys.3.61}. If all
chemical bond formations, electron hopping and effects leading to band-energy
dispersion in a solid take place outside $r_c$, one can replace the
all-electron potential around each atom position of a solid, with a lattice of
pseudopotentials. Clearly, this can be extended to molecular and cluster
entities.

\begin{figure} \centering 
\includegraphics[width=\linewidth]{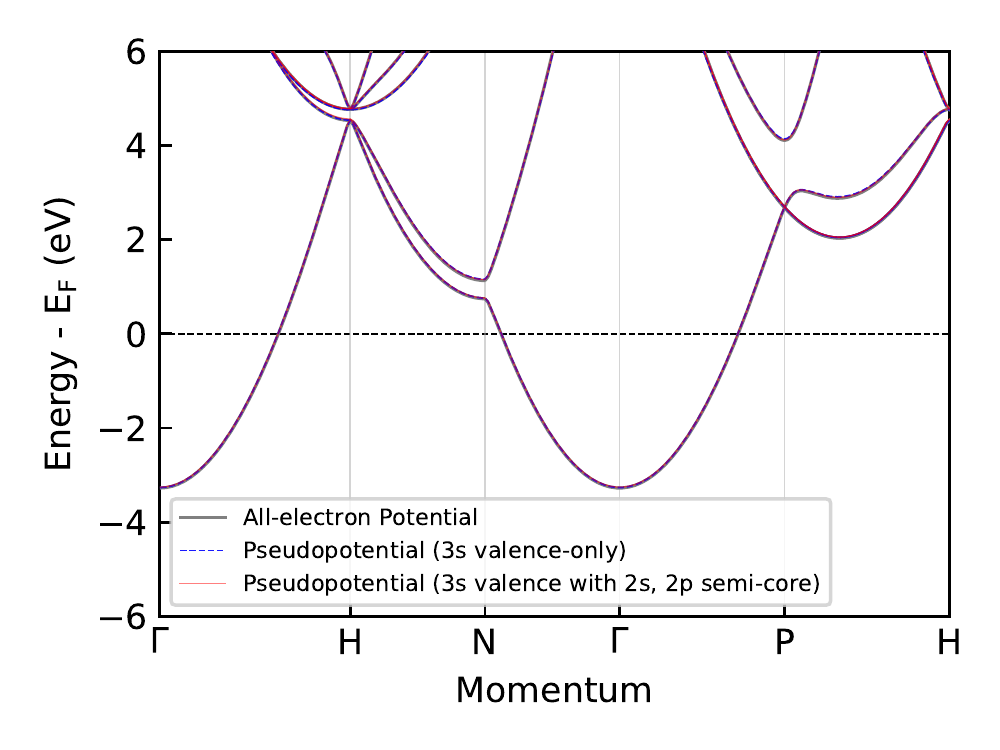}
	\caption{Comparison of the DFT energy dispersion for bcc Na
	 within the local density approximation (LDA) using experimental equilibrium
	unit-cell volume, calculated using the full-potential all-electron 
	electronic structure method	with RSPt (grey) and the pseudopotential method with Quantum Espresso. Two atomic pseudopotentials are used, with (red) and without (blue) the semi-core 2$s$ state. The energy levels are plotted relative to the Fermi level, $E_F$. These calculations use the same convergence parameters as in Fig.~\ref{Fig:Eigenvalues}.  }
\label{Fig:RsptvsQE} 
\end{figure}

In a recent study~\cite{Science.351.aad3000}, \textit{ab initio} pseudopotential
electronic structure results were found to be in good agreement
with those  computed using all-electron theory. Indeed,   the comparison in 
Ref.~\cite{Science.351.aad3000} was made for 71 elements, and the 
agreement of the  all-electron and pseudopotential results,  for 
elements with very different types of chemical bonding, supports the use
of the computationally more efficient pseudopotential method. 
An example of comparison of the energy dispersion calculated using the
all-electron and  the standard nonlocal pseudopotential methods is shown in Fig.~\ref{Fig:RsptvsQE}. The figure illustrates the energy bands of bcc Na at
the experimental unit-cell volume $V_0$~\cite{Barrett1956} (corresponding to the lattice constant = 4.2250~{\AA}). It is
clear from the figure that pseudopotential- and all-electron electronic structure theory can produce very similar band dispersion if the pseudopotential is properly constructed.

For  a pseudopotential to have general applicability, it is important that it
is transferable~\cite{Goedecker1992, Filippetti1995} to different chemical
environments, e.g., works for a molecule, or a material that forms covalent or ionic
bonds, under ambient or high pressure, or even for a crystal surface. The
transferability of a pseudopotential characterizes the accuracy with which the
pseudopotential reproduces the effect of the all-electron potential in different chemical
environments. One way to test for the transferability of a pseudopotential to a
different chemical environment is by comparing the  calculated Kohn-Sham energy
eigenvalues with the self-consistent all-electron results~\cite{Goedecker1992}.
This raises the question of whether one can apply the normal protocols for generating
pseudopotentials, but instead of using the atomic state as reference electronic configuration take an environment
that more closely resembles the native conditions where the pseudopotential is
supposed to be used, be it in a solid or molecular state. We will refer to the standard  pseudopotentials derived using the free isolated atom as the reference configuration as an atomic pseudopotential. In this work, we also use the solid state as a reference
configuration. We refer to such pseudopotentials as \textit{in-situ} pseudopotentals, and will demonstrate as a proof-of-concept in this paper that these  
pseudopotentials can reproduce all-electron results to very high accuracy.

The advantage of an \textit{in-situ} pseudopotental is that it is tailored to the specific chemical environment of the
material (e.g. under high compression) and,
as a result, it can in general be used as an expedient, accurate and
computationally inexpensive tool to analyze electronic structures of complex
systems, e.g. as discussed in Ref~\cite{Sarma_1989}. 
In addition, the use of the solid-state environment to generate \textit{in-situ} pseudopotentials is
motivated by the fact that the scattering properties of a pseudopotential
constructed for an isolated atom might be  different from  those of the  same
atom  placed in a material.  This is particularly of concern, when the
environment in the solid is drastically different from that of an  atom, e.g., when neighboring atoms in an ionic bonded material cause large charge
transfer effects that affect the multiple scattering properties. Similarly, it can
be troublesome to use an atomic generated pseudopotential evaluated at ambient
conditions, for a solid state calculation of the electronic structure of a
material under extreme compression. 

In this paper, we outline the critical steps to generate \textit{in-situ} pseudopotentials, and calculate for bcc Na the band dispersion using an \textit{in-situ} pseudopotential generated from an all-electron reference state obtained from the full-potential linear muffin-tin orbitals method (RSPt software~\cite{Wills2010}). This result is then compared with the band structure obtained from two different atomic (i.e., generated from an atomic reference state), scalar-relativistic and norm-conserving pseudopotentials using the Quantum Espresso software~\cite{Giannozzi2009}. The main difference is that the first atomic pseudopotential (i) contains only the valence 3$s$ state of Na (i.e., valence-only) while the second (ii) contains not only the valence 3$s$ state, but also the $2s$ and $2p$ semi-core states. The former~\cite{note1} pseudopotential is a Troullier-Martins~\cite{Troullier1991} pseudopotential generated using FHI98PP~\cite{Fuchs1999} and includes nonlinear core correction~\cite{Louie1982}. It has only one Kleinman-Bylander-Vanderbilt~\cite{PhysRevB.41.7892, PhysRevLett.48.1425} projector per angular channel. The latter pseudopotential is an optimized norm-conserving Vanderbilt pseudopotential (ONCVPSP)~\cite{Hamann2013} obtained from the \textsc{PseudoDojo} project~\cite{PseudoDojo2018}. It does not use nonlinear core correction and has two projectors per angular channel. All pseudopotential calculation with Qunatum Espresso uses a kinetic-energy cutoff of 100~Ry for the plane wave basis expansion, a $\mathbf{k}$-grid of $24 \times 24 \times 24$ and LDA~\cite{PerdewWang1992} for the self-consistent DFT calculation. 

\section{Method} \label{Sec:Method} 

Simply described, the method to generate \textit{in-situ} pseudopotentials can be divided
into three distinct steps.  (1) Calculate the eigenstates of the Kohn-Sham
equations of the solid using an all-electron method. (2) Construct the pseudowavefunction by modifying the corresponding valence state to remove any nodes in the core region while exactly preserving the wavefunction in
the interstitial region. (3) Generate a pseudopotential that gives
rise to this pseudowavefunction. We begin by describing step (3) in
Sec.~\ref{Sec:InvPseudoPotProb} before moving on to the more technical details
of step (2) in Sec.~\ref{Sec:WfnMod}.

\subsection{The inverse pseudopotential problem} \label{Sec:InvPseudoPotProb}
For the purpose of this section, we first assume that the pseudowavefunction,
$\tilde{\Psi}_{\mathbf{k}n}$, and its energy eigenvalue,
$\tilde{\epsilon}_{\mathbf{k}n}$, for each $\mathbf{k}$ point in the Brillouin zone and band $n$, are already known. 
Next, consider the eigenvalue equation 
\begin{align} \label{Eq:SE}
H_{pp}\tilde{\Psi}_{\mathbf{k}n} =& \tilde{\epsilon}_{\mathbf{k}n}\tilde{\Psi}_{\mathbf{k}n}, 
\end{align} 
where $\mathbf{k}$ is crystal momentum restricted to the first Brillouin zone, $n$ is
a band index for the given $\mathbf{k}$, and
\begin{align} 
\label{Eq:PseudoPot1}
H_{pp} =& -\nabla^2 + V_{pp}(\mathbf{r})\\
\label{Eq:PseudoPot} 
=& -\nabla^2 + \sum_{\mathbf{G}}V_{\mathbf{G}}e^{i\mathbf{G}\cdot\mathbf{r}},\\
\label{Eq:PseudoWfn} 
\tilde{\Psi}_{\mathbf{k}n}(\mathbf{r}) =& \sum_{\mathbf{G}}\tilde{\CalD}_{\mathbf{G}}^{\mathbf{k}n}e^{i(\mathbf{G}+\mathbf{k})\cdot\mathbf{r}}.
\end{align} 
Here, $\mathbf{G}$ runs over the reciprocal lattice vectors, and
$V_{\mathbf{G}}$ and $\tilde{\CalD}_{\mathbf{G}}^{\mathbf{k}n}$ are,
respectively, the plane wave expansion coefficients for the pseudopotential
$V_{pp}$ and pseudowavefunction, $\tilde{\Psi}_{\mathbf{k}n}$. The 
$\tilde{\Psi}_{\mathbf{k}n}$ is known as the pseudowavefunction as it is the
solution of a Hamiltonian involving a pseudopotential. In 
Eqs.~~\ref{Eq:PseudoPot1} and \ref{Eq:PseudoPot}, we have used
Rydberg atomic units.  Multiplying Eq.~\ref{Eq:SE} by
$e^{-i(\mathbf{G^\prime+k})\cdot\mathbf{r}}$ and integrating over $\mathbf{r}$, we
obtain the expression, 
\begin{align} 
\label{Eq:EigenvalEq}
{\left(\mathbf{G}+\mathbf{k}\right)^2}\tilde{\CalD}_{\mathbf{G}}^{\mathbf{k}n} +
\sum_{\mathbf{G}^\prime}V_{\mathbf{G}^\prime}\tilde{\CalD}_{\mathbf{G}-\mathbf{G}^\prime}^{\mathbf{k}n}
=&\;\tilde{\epsilon}_{\mathbf{k}n}\tilde{\CalD}_{\mathbf{G}}^{\mathbf{k}n}.
\end{align}

In practice, the sum over $\mathbf{G^\prime}$ in the plane wave expansion of the
pseudopotential and pseudowavefunction is truncated for a finite number ($N$), of
coefficients.  Therefore, for a given eigenstate, Eq.~\ref{Eq:EigenvalEq}
corresponds to $N$ linear equations with $N$ unknowns, which in matrix form can
be written as (note that $\mathbf{k}$ and $n$ are state labels, not matrix
indices) 
\begin{align} \label{Eq:PseudoPotEqnMat}
\mathbf{M}^{\mathbf{k}n}\mathbf{v} = \mathbf{u}^{\mathbf{k}n}, 
\end{align}
where 
\begin{align} \label{Eq:CrystalPot} 
\left[\mathbf{v}\right]_{\mathbf{G}} =& \;V_{\mathbf{G}},\\ \label{Eq:PseudoWfnCoeff}
\left[\mathbf{u}^{\mathbf{k}n}\right]_{\mathbf{G}} =& \left(\tilde{\epsilon}_{\mathbf{k}n} - {(\mathbf{G} +
\mathbf{k})^2}\right)\tilde{\CalD}_{\mathbf{G}}^{\mathbf{k}n},\\
\left[\mathbf{M}^{\mathbf{k}n}\right]_{\mathbf{G},\mathbf{G}'} =&
\;\tilde{\CalD}_{\mathbf{G}-\mathbf{G}'}^{\mathbf{k}n}.  
\end{align}
The matrix $\mathbf{M}^{\mathbf{k}n}$ is a blocked Toeplitz matrix. The linear system of equations in Eq.~\ref{Eq:PseudoPotEqnMat} can therefore be solved by a blocked Levinson algorithm, which formally produce the pseudopotential 
\begin{align} 
\label{Eq:PseudoPotSol} \mathbf{v} =&
\left(\mathbf{M}^{\mathbf{k}n}\right)^{-1}\mathbf{u}^{\mathbf{k}n}.
\end{align} 
While this pseudopotential is only constructed to exactly reproduce the eigenvalue
$\tilde{\epsilon}_{\mathbf{k}n}$ for a
particular $\mathbf{k}$, we will see in Sec.~\ref{Sec:Results} that
the procedure gives a pseudopotential that gives satisfactory results
for eigenvalues throughout the Brillouin zone, i.e. also for eigenvalues
calculated at $\mathbf{k}^{\prime} \ne \mathbf{k}$.

\subsection{Choice of wavefunction and practical implementation}
\label{Sec:WfnMod} While the method described in
Sec.~\ref{Sec:InvPseudoPotProb} is straightforward, the difficulty lies in
generating an appropriate wavefunction $\tilde{\Psi}_{\mathbf{k}n}$ to use as
input. In principle one can calculate it by first solving an all-electron
electronic structure problem, thereby obtaining the true energy eigenvalue and
wavefunction. Hence, by finding a solution to 
\begin{align} \label{eq:AE_SE}
H_{AE}{\Psi}_{\mathbf{k}n}(\mathbf{r}) =& {\epsilon}_{\mathbf{k}n}{\Psi}_{\mathbf{k}n}(\mathbf{r}),
\end{align} 
where $H_{AE}$ is the all-electron Hamiltonian, one could in
principle solve Eqs.~\ref{Eq:CrystalPot} - \ref{Eq:PseudoPotSol}, by setting
$\tilde{\Psi}_{\mathbf{k}n}$ equal to the true all-electron wavefunction,
${\Psi}_{\mathbf{k}n}$, and by identifying $\tilde{\epsilon}_{\mathbf{k}n}$
with ${\epsilon_{\mathbf{k}n}}$. In this way one is guaranteed that the
pseudopotential that comes out of Eq.~\ref{Eq:PseudoPotSol} gives the same
eigenvalue and wavefunction as the all-electron Hamiltonian. A
pseudopotential generated at one particular $\mathbf{k}$-point, e.g. the
$\Gamma$-point, can then be used in Eqs.~\ref{Eq:SE}~-~\ref{Eq:PseudoPot}, to
calculate eigenvalues and eigenstates throughout the Brillouin zone. One can
also envision using this pseudopotential for other, but similar, conditions,
e.g. a crystal at compressed volumes compared to the condition where the
pseudopotential is originally calculated from. In this approach, the valence
states generated by an all-electron calculation are expected to have nodes in
the core region, which will require a very large number of the basis vectors to
converge the calculation if the plane wave basis set is used
(Eq.~\ref{Eq:PseudoWfn}).
A solution to this problem is to replace the fast oscillating part of
${\Psi}_{\mathbf{k}n}$, that is primarily close to an atomic nucleus, with a
smooth pseudowavefunction, $\tilde{\Psi}_{\mathbf{k}n}$, that is nodeless,
while still keeping $\tilde{\epsilon}_{\mathbf{k}n} =
{\epsilon_{\mathbf{k}n}}$. This is the usual way of pseudopotential theory,
albeit we propose here to do it using its native solid state as the reference
electronic configuration, and not the free atom. 

The description that follows aims at describing how a smooth, nodeless
pseudowavefunction can be evaluated from an all-electron wavefunction that is
obtained from a full-potential linear muffin-tin orbitals method, as
implemented in the RSPt package~\cite{Wills2010}. We start with the general
approach of writing the all-electron wavefunction as a linear combination of
known basis functions, namely, the linear muffin-tin orbitals (LMTOs),

\begin{align} \label{eq:LMTO_basis} {\Psi}_{\mathbf{k}n}(\mathbf{r})=\sum_\Lambda
	{\CalC}_\Lambda^{\mathbf{k}n} \phi_\Lambda^{\mathbf{k}}(\mathbf{r}), 
\end{align} where
$\phi_\Lambda^{\mathbf{k}}$ are the LMTOs introduced by
Andersen~\cite{PhysRevB.12.3060}, where  the $\Lambda$ index groups many indices, 
such as the tail energy of the basis function, angular momenta and type of atomic species. 
We emphasize that $\mathbf{k}n$ are state
labels, not indices. The LMTOs are defined with respect to two regions: the
muffin-tin and the interstitial regions. In the latter, the LMTO basis function
is either a Hankel or a Neumann function, depending on choice of kinetic energy
of this basis function. For practical reasons, in the RSPt package~\cite{Wills2010}, the wavefunction in the
interstitial region is calculated as an exact Fourier series. This is done by
extending a Hankel or Neumann function from the interstitial region into the
muffin-tin sphere with an analytic smooth function, 
for fast converged Fourier series  expansion.
This means that the interstitial basis function is
defined over all space, and matrix elements of e.g. the Hamiltonian or the
Bloch wavefunction overlap is truncated inside the muffin-tins using  a step function. 

Following Ref.~\cite{Wills2010}, we define a pseudo basis function, $\tilde
\phi_\Lambda^{\mathbf{k}}$, as the Fourier-transformed Hankel or Neumann function,
that in the interstitial region is identical to the all-electron LMTO basis function:

\begin{align} \label{Eq:AE_PseudoPot} \tilde \phi_\Lambda^{\mathbf{k}}(\mathbf{r}) 
=\sum_{\mathbf{G}} 
	\tilde A_{\Lambda\mathbf{G}}^{\mathbf{k}} e^{i(\mathbf{G}+\mathbf{k}) \cdot\mathbf{r}}, 
\end{align} 
where $\tilde A_{\Lambda\mathbf{G}}^{\mathbf{k}}$ are the Fourier coefficients of this basis
function. This function can now be used for the expression of the Fourier
series in Eqs.~\ref{Eq:SE}~-~\ref{Eq:PseudoWfn}, that should be defined over
all space, i.e. including the muffin-tin region. We start by considering the
all-electron wavefunction, in Eq.~\ref{eq:AE_SE}, that in the interstitial can also be
expressed in terms of the Fourier series, in Eq.~\ref{Eq:AE_PseudoPot}. This
can be done by replacing $\phi_\Lambda^{\mathbf{k}n}$ in Eq.~\ref{eq:LMTO_basis} with
$\tilde \phi_\Lambda^{\mathbf{k}n}$. By construction, this replacement does not
influence the wavefunction in the interstitial region. However, it drastically
modifies the wavefunction in the muffin-tin region, since the part of the
muffin-tin orbital that is defined in the muffin-tin sphere, where in general
the radial component has many nodes, is replaced by a smooth function. For this
reason we distinguish the true wavefunction, Eq.~\ref{eq:LMTO_basis}, from a
pseudowavefunction,

\begin{align} \label{Eq:AE_PseudoWfn} \tilde {\Psi}_{\mathbf{k}n}(\mathbf{r})= \sum_\Lambda
{\CalC}_\Lambda^{\mathbf{k}n} \tilde \phi_\Lambda^{\mathbf{k}}(\mathbf{r}).  
\end{align} 
Notice that the
expansion coefficients in Eqs.~\ref{eq:LMTO_basis} and \ref{Eq:AE_PseudoWfn},
${\CalC}_\Lambda^{\mathbf{k}n}$, should be the same. Writing out explicitly the form of
$\tilde \phi_\Lambda^{\mathbf{k}n}$ we can express the pseudowavefunction as

\begin{align} \tilde {\Psi}_{\mathbf{k}n}(\mathbf{r})=\sum_\Lambda {\CalC}_\Lambda^{\mathbf{k}n}
\sum_{\mathbf{G}} \tilde{\CalA}_{\Lambda\mathbf{G}}^{\mathbf{k}} e^{i(\mathbf{G}+\mathbf{k}) \cdot\mathbf{r}}.
\end{align} 
Using this equation, the coefficients $\tilde{\CalD}$ of Eq.~\ref{Eq:PseudoWfn} will be given by


\begin{align} \label{Eq:Final_AE_PP} 
	\tilde {\CalD}^{\mathbf{k}n}= \sum_\Lambda {\CalC}_\Lambda^{\mathbf{k}n} 
	\tilde{\CalA}_{\Lambda\mathbf{G}}^{\mathbf{k}}.
\end{align} 

The  form given by the latter equation should be  used in the pseudowavefunction given   in Eq.~\ref{Eq:PseudoWfn} to calculate the corresponding
pseudopotential, by following
Eqs.~\ref{Eq:PseudoPotEqnMat}~-~\ref{Eq:PseudoPotSol}.  A practical way to
evaluate a pseudopotential with this method is to first perform a normal
all-electron calculation to obtain ${\CalC}_\Lambda^{\mathbf{k}n}$ coefficients (and
the eigenvalue, $\epsilon_{\mathbf{k}n}$). In this process, the Fourier
coefficients of the pseudo-basis function are kept (from Eqs.~\ref{Eq:AE_PseudoPot} and \ref{Eq:Final_AE_PP}), which enables an evaluation
of Eq.~\ref{Eq:PseudoWfn}. The so-obtained pseudowavefunction and
eigenvalue are used in Eqs.~\ref{Eq:PseudoPotEqnMat}~-~\ref{Eq:PseudoPotSol},
to obtain the required pseudopotential.

Although the description above seems straight forward, we note that these
modifications are done in the full-potential method of Ref.~\cite{Wills2010},
independently of whether a pseudopotential is to be extracted or not. They are
in line with the aims of the pseudopotential approach, but are strictly
speaking related to the computational benefits associated with having fewer
coefficients in the Fourier expansion. To be useful for a pseudopotential
approach, we also need to ensure that low-lying `ghost states' do not appear.
To understand the `ghost state' problem, consider applying
Eq.~\ref{Eq:PseudoPotSol} immediately to the unmodified valence state.  By
construction, the resulting pseudopotential gives rise to a Hamiltonian that
contains this eigenstate.  However, the full Hilbert space also contains
smoother states and these tend to have lower energy.  This is not surprising
since states with lower energy do exist in the original problem, namely the
core states.  The purpose of the pseudopotential approach is to generate an
effective Hamiltonian for which the valence states are the low energy states
and it is therefore essential to remove radial nodes in the wavefunction.
While the procedure outlined above does reduce the number of radial nodes, it
is not constructed to guarantee an absence of nodes. 
It also does not guarantee norm conservation. In this work, we are
focused on constructing norm-conserving pseudopotentials (even though this
constraint of norm-conservation can be relaxed in future works, similar to that
in the ultrasoft-pseudopotential~\cite{PhysRevB.41.7892, PhysRevB.41.5414} and
PAW methods~\cite{PhysRevB.50.17953, PhysRevB.59.1758}, at the expense of a
more complex mathematical representation, versus the simpler
representation of norm-conserving pseudopotentials~\cite{Hamann1989, Troullier1991}).

For these reasons we modify the pseudowavefunction further to make it nodeless 
and ensure norm conservation.  It is the aggregate of these
modifications to the pseudowavefunction that are compensated for through the
pseudopotential.  Only by modifying the wavefunction in the core region is it
possible to preserve the chemical properties that emerge from the
pseudopotential, since chemical bonding is mainly determined by the
wavefunction overlap in the interstitial region between atoms.  The
expression we arrived at for the nodeless, normalized pseudowavefunction is
\begin{align} 
\label{Eq:CoreReplacement} \hat{\Psi}_{\mathbf{k}n}(r, \theta,
\varphi) = c(r)\tilde{\Psi}_{\mathbf{k}n}(r, \theta, \varphi) + N(1 -
c(\mathbf{r}))f(r, \theta, \varphi), 
\end{align} 
where $f(r, \theta, \phi)$ is
a smooth function with the same angular dependence as
$\tilde{\Psi}_{\mathbf{k}n}(r, \theta, \varphi)$ and $c(\mathbf{r})$ smoothly
interpolates between $\tilde{\Psi}^{\mathbf{k}n}(r, \theta, \varphi)$ in the
interstitial region and $f(r, \theta, \varphi)$ in the muffin-tin region.  For
convenience, we have also factored out a constant $N$ from $f(r, \theta,
\varphi)$ that will be used to ensure that the pseudowavefunction is normalized. Note that Eq.~\ref{Eq:CoreReplacement} is general, in sense that it can be
applied to an all-electron valence state, although we for technical reasons
apply it to the pseudowavefunction obtained from
Eq.~\ref{Eq:Final_AE_PP}, $\tilde{\Psi}^{\mathbf{k}n}$ .

In this manuscript we provide a proof-of-principle demonstration of the method
for the energy dispersion of the Na $3s$-band. Adapted for this state we
choose 
\begin{align} 
\label{Eq:CoreReplacement2}
    c(r) =& 
    \left\{\begin{array}{cc} 
             0        & |r| < R_0,\\
             1        & |r| > R_1,\\
             \left.3x^2 - 2x^3\right|_{x=\frac{r - R_0}{R_1 - R_0}}
                      & \textrm{otherwise}. \\ 
           \end{array}
   \right.
\end{align} 
The node-free, normalized pseudowavefunction is then defined, with a suitable choice of $f(r, \theta, \phi)$ in Eq.~\ref{Eq:CoreReplacement}. We return to appropriate choices of this function below. First we remark that 
in Eq.~\ref{Eq:CoreReplacement2}, $R_0$ and $R_1$ are
chosen so as to obtain a smooth interpolation without nodes. In general, $R_0$
should be larger than the radius of the outermost radial node, and we have $R_0
< R_1 < R_{MT}$, where $R_{MT}$ is the muffin-tin radius. Returning now to the choice of $f(r, \theta, \phi)$, we have in the numerical examples presented below focused on the valence state of Na, which is dominated by the $3s$ state. This function has no angular component and it is sufficient to use $f(r, \theta, \phi)=f(r)$. We have made two choices for $f(r)$; a constant value of 1 and   
a polynomial of degree 15. In the latter choice, we determined the expansion coefficients through a least square fit of a pseudowavefunction obtained from a Quantum Espresso calculation using the valence-only pseudopotential (this is shown as a green line in Fig. \ref{Fig:Eigenvalues}a). Below we will compare the results for both choices of $f(r)$. 

The constant, $N$, in Eq.~\ref{Eq:CoreReplacement} is determined by requiring that
$\hat{\Psi}_{\mathbf{k}n}$ is normalized to 1, 
\begin{widetext}
\begin{align}
	\langle\hat{\Psi}_{\mathbf{k}n}|\hat{\Psi}_{\mathbf{k}n} \rangle =
	\langle c\tilde{\Psi}_{\mathbf{k}n}|c\tilde{\Psi}_{\mathbf{k}n}\rangle
	+ N^2\langle(1 - c)f|(1-c)f\rangle + 2N\textrm{Re}\left(\langle
	c\tilde{\Psi}_{\mathbf{k}n}|(1-c)f\rangle\right) = 1.  
\end{align} 
\end{widetext}
This is solved by, 
\begin{align} N =& -\frac{B}{C}
\pm\frac{\sqrt{\left(1 - A\right)B + C^2}}{B}, \end{align} where,
\begin{align} A =& \langle
	c\tilde{\Psi}_{\mathbf{k}n}|c\tilde{\Psi}_{\mathbf{k}n}\rangle,\\
	B =& \langle(1-c)f|(1-c)f\rangle,\\ C =& \textrm{Re}\left(\langle
	c\tilde{\Psi}_{\mathbf{k}n}|(1-c)f\rangle\right).  
\end{align}

\section{Results} \label{Sec:Results} 
\begin{figure*} \centering
\includegraphics[width=\linewidth]{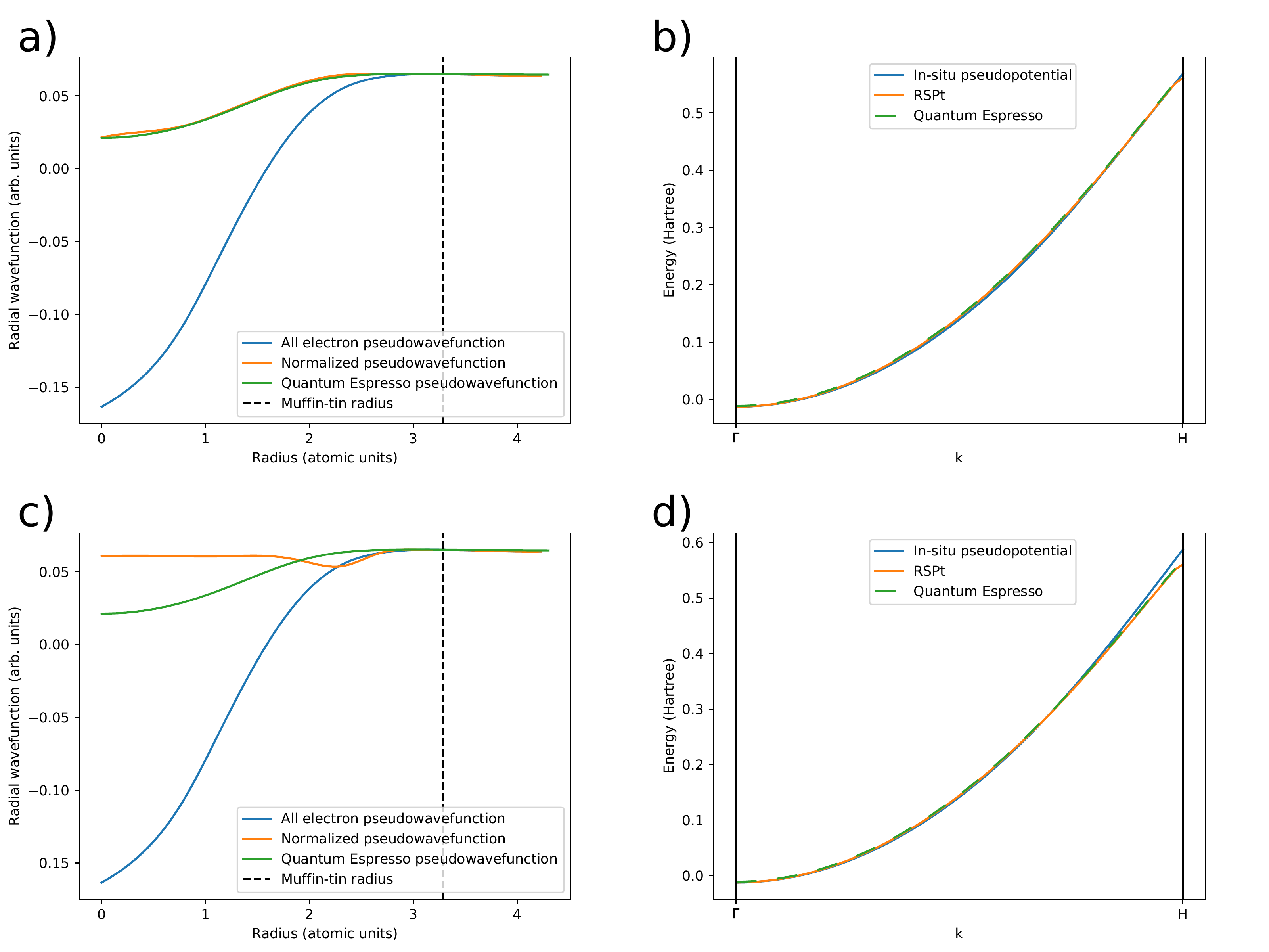} 
\caption{(a and c) Calculated radial component of the pseudowavefunction $\tilde{\Psi}_{\mathbf{k}n}(r, \theta, \varphi)$ (blue, for definition see text),  calculated radial part of the modified pseudowavefunction $\hat{\Psi}_{\mathbf{k}n}(r, \theta, \varphi)$ (orange, for definition see text), and radial pseudowavefunction calculated with Quantum Espresso using a valence-only atomic pseudopotential (green). (b and d) Calculated energy dispersion of the valence band states of bcc Na along the $\Gamma - H$ high symmetry direction of the first Brillouin zone. Three types of methods are used to calculate the energy bands: the all-electron full potential method (orange), the \textit{in-situ} pseudopotential method as described of this paper (blue) and the pseudopotential method using a valence-only atomic pseudopotential (green). Panels (a) and (b) show results for the choice of $f(r)$ being a polynomial of degree 15 that is least-square fitted to the radial pseudowavefunction calculated with Quantum Espresso using a valence-only pseudopotential (using $R_0 = 0.75R_{MT}$ and $R_1 = 0.9R_{MT}$). Panels (c) and (d) show results for a choice of  $f(r) = 1$ (using $R_0 = 0.55R_{MT}$, and $R_1 = 0.75R_{MT}$. The muffin-tin radius, $R_{MT}=3.285$~\AA, is denoted using dashed lines in all panels). In all these calculations, a $11\times 11\times 11$ mesh is used for the Fourier expansion of the pseudopotential.}
\label{Fig:Eigenvalues} \end{figure*}

In Fig.~\ref{Fig:Eigenvalues}, we
compare the radial components of $\hat{\Psi}_{\mathbf{k}n}(r, \theta, \varphi)$
and $\tilde{\Psi}_{\mathbf{k}n}(r, \theta, \varphi)$ for the lowest eigenvalue
of bcc Na, obtained at the $\Gamma$ point. Note that we show in the figure results of pseudowavefunction and band dispersion for two choices of $f(r)$ in Eq.~\ref{Eq:CoreReplacement}: a constant and a 15-degree polynomial.
Included for reference is also a pseudowavefunction obtained from a Quantum Espresso calculation (from which one choice of $f(r)$ was obtained). In this calculation, we set the
muffin-tin radius, $R_{MT} = 3.285$~a.u., which is approximately 95\% of the
touching radius between two nearest Na atoms, and use $R_0 = 0.55R_{MT}$ and
$R_1=0.75R_{MT}$ for the constant $f(r)$ and $R_0 = 0.75R_{MT}$ and
$R_1=0.9R_{MT}$ for the polynomial. The all-electron calculation that is used to evaluate the
pseudowavefunction uses the local density approximation
(LDA)~\cite{Hedin_1971, Barth_1972}. As LMTO basis vectors, we use three 3$s$
orbitals, three 3$p$ orbitals and two 3$d$-orbitals in the muffin-tin spheres.
In the interstitial region, the tails have  kinetic energies of 0.3~Ry (for
3$s$-, 3$p$- and 3$d$-orbitals), -2.3~Ry (for 3$s$-, 3$p$- and 3$d$-orbitals)
and -1.5~Ry (for 3$s$- and 3$p$-orbitals). The number of $\mathbf{k}$-points
used to converge the results were $12\times 12\times 12$. As in
Fig.~\ref{Fig:RsptvsQE}, the experimental~\cite{Barrett1956} lattice parameter
of 4.225~{\AA} was used.

It is clear from Fig.~\ref{Fig:Eigenvalues} (a and c)
that
$\hat{\Psi}_{\mathbf{k}n}(r, \theta, \varphi)$ and
$\tilde{\Psi}_{\mathbf{k}n}(r, \theta, \varphi)$  are identical in the
interstitial region. These wavefunctions also coincide with the full, true
all-electron wavefunction in the interstitial region (data not shown).  
From a
detailed inspection of the radial components of $\hat{\Psi}_{\mathbf{k}n}(r,
\theta, \varphi)$ and $\tilde{\Psi}_{\mathbf{k}n}(r, \theta, \varphi)$ we note
that the latter has a single node, as opposed to the two nodes expected from a 3$s$
state. The single node of the otherwise rather soft behavior of
$\tilde{\Psi}_{\mathbf{k}n}(r, \theta, \varphi)$, has to do with how the
full-potential method of Ref.~\cite{Wills2010} represents the basis functions
in the interstitial, in particular as a Fourier series (see Eqs.~6.38~-~6.42 of
Ref.~\cite{Wills2010}).  In order to obtain a
pseudopotential that is as smooth as possible from Eqs.~\ref{Eq:PseudoPotEqnMat}~-~\ref{Eq:PseudoPotSol}, we
have made use of $\hat{\Psi}_{\mathbf{k}n}(r, \theta, \varphi)$ (in a Fourier
representation) instead of $\tilde{\Psi}_{\mathbf{k}n}(r, \theta, \varphi)$,
since the former pseudowavefunction is by construction node-less inside the
muffin-tin sphere (see Fig.~\ref{Fig:Eigenvalues}a).  This choice leads
to a much smoother pseudowavefunction, that may be expressed with a minimum
number of Fourier components. For this reason we have used the Fourier
representation of $\hat{\Psi}_{\mathbf{k}n}(r, \theta, \varphi)$ in all the
steps outlined in Eqs.~\ref{Eq:SE}~-~\ref{Eq:PseudoPotSol}, discussed in
Sec.~\ref{Sec:Method}.

In Fig.~\ref{Fig:Eigenvalues} (a and c), we note that depending on choice of $f(r)$, the behaviour of $\hat{\Psi}_{\mathbf{k}n}(r, \theta, \varphi)$ inside the muffin-tin region is different. For the choice of a 15-degree polynomial for $f(r)$, $\hat{\Psi}_{\mathbf{k}n}(r, \theta, \varphi)$ is by construction similar to the function obtained from a calculation based on 
 Quatum Espresso (see Fig.~\ref{Fig:Eigenvalues}a). Although the behavior in the core region is explicitly constructed, a good match is not guaranteed from the outset. The freedom provided through the normalization constant $N$, and the fact that the two regions (interstitial and muffin-tin) are stitched together with the help of the interpolation function $c(r)$ rather than being matched at the muffin-tin boundary, means the two functions are allowed to differ. The near-perfect match in spite of this is a reassurance of the soundness of the interpolation procedure.


After having calculated a pseudopotential, using
Eqs.~\ref{Eq:SE}~-~\ref{Eq:PseudoPotSol} in combination with
$\hat{\Psi}_{\mathbf{k}n}(r, \theta, \varphi)$, we calculate eigenvalues and
eigenvectors from Eq.~\ref{Eq:EigenvalEq}. This represents therefore the
solution to a local pseudopotential, and the calculation was done using 1331
plane-wave components in the expansion of the wavefunction.  In
Fig.~\ref{Fig:Eigenvalues} (b and d) we show the resulting energy dispersion along the
high-symmetry line, $\Gamma - H$, of the first Brillouin zone, for two different \textit{in-situ} pseudopotentials, one from a choice of $f(r)$ as a constant and one from a choice of $f(r)$ being a 15-degree polynomial. The two different \textit{in-situ} pseudopotential results are 
compared to the energy dispersion of an all-electron full-potential electronic
structure method (see Fig.~\ref{Fig:Eigenvalues} b and d) . We first note that the whole methodology described above is
designed to yield the same eigenvalue and wavefunction in the interstitial
region at the $\Gamma$ point. Hence, it is gratifying that for the $\Gamma$ point the eigenvalues from the
\textit{in-situ} pseudopotential method and the all-electron full-potential method differ only
in the sixth significant digit, irrespective of choice of $f(r)$ in Eq.~\ref{Eq:CoreReplacement2}. Furthermore, the energy dispersion is seen to
agree very well between \textit{in-situ} pseudopotential theory and all-electron theory throughout the Brillouin
zone. 
When comparing the two \textit{in-situ} pseudopotentials (evaluated from constant and polynomial choice of $f(r)$), we note that 
the agreement in band dispersion between all-electron theory and any \textit{in-situ} pseudopotential theory is surprisingly good. This holds true even when we set $f(r)=1$. The latter has a pseudowavefunction that in the core region differs significantly from a traditional behaviour (e.g. as seen from the results obtained from the Quantum Espresso calculation). The poorer choice of $f(r)=1$ nevertheless results in an \textit{in-situ} pseudopotential that reproduces all-electron results throughout most of the Brillouin zone, demonstrating the robustness of our approach.
The largest difference for the eigenvalues is observed at the zone boundary, which is not
unexpected, considering that these states have crystal momentum farthest away from that state where the \textit{in-situ} pseudopotential was calculated (the $\Gamma$ point).

\begin{figure*} \centering
\includegraphics[width=\linewidth]{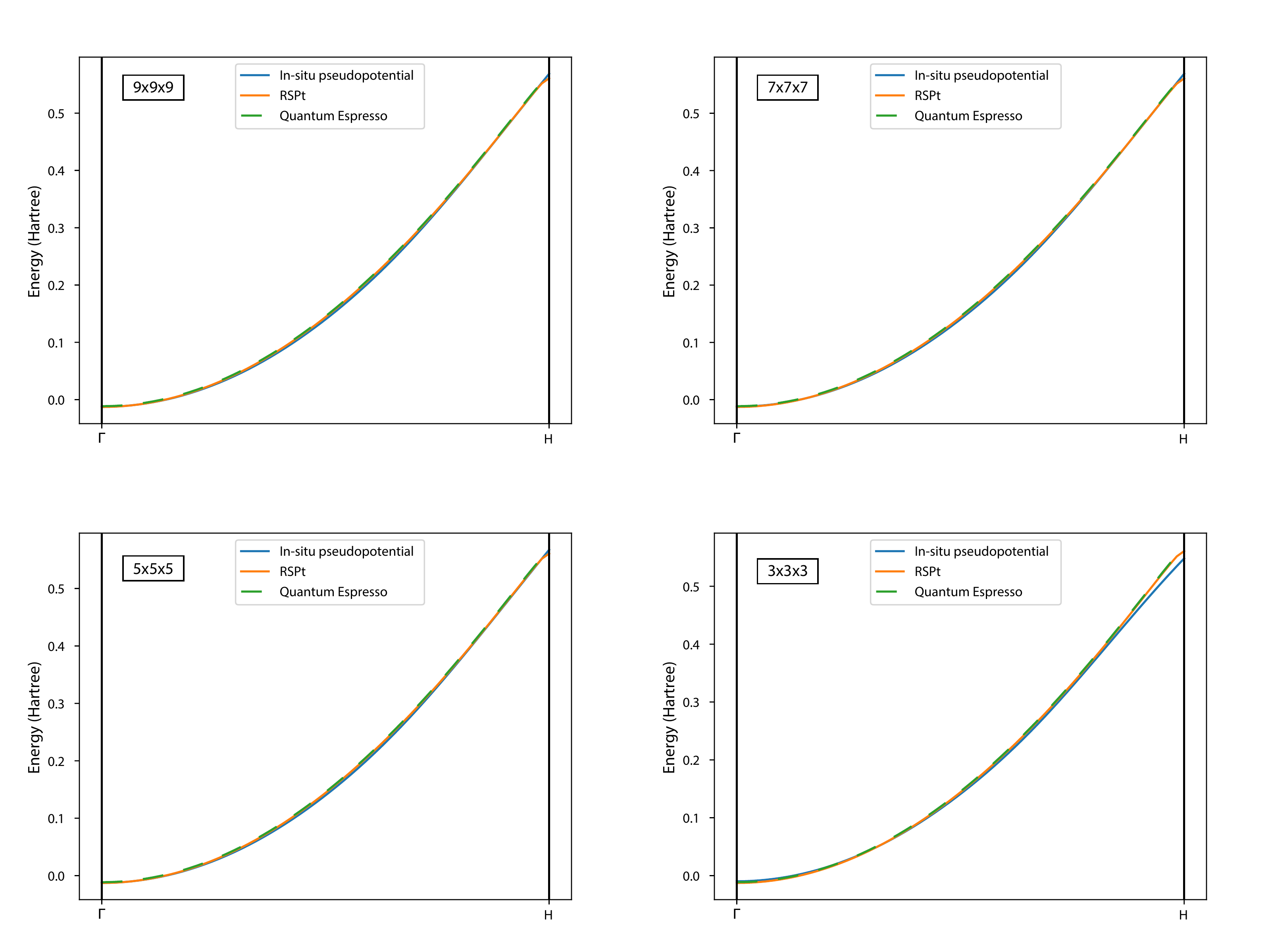}
\caption{Calculated energy dispersion of the valence band states of bcc Na, along the
$\Gamma - H$ high symmetry direction of the first Brillouin zone. It is similar to the plot in Panel (b) of Fig.~\ref{Fig:Eigenvalues}, except that the Fourier expansion of the psuedopotential is truncated to a smaller $N\times N\times N$ mesh as specified in the inset of each panel. As in Panel (b) of Fig.~\ref{Fig:Eigenvalues}, three types of methods are used to calculate the energy bands: the all-electron full potential method (orange), the \textit{in-situ} pseudopotential method as described of this paper (blue) and the pseudopotential method using a valence-only atomic pseudopotential (green). Here, $f(r)$ is the 15-degree polynomial least-square fitted to the radial pseudowavefunction calculated with Quantum Espresso using a
valence-only pseudopotential (using $R_0 = 0.75R_{MT}$ and $R_1 = 0.9R_{MT}$)}
\label{Fig:EigenvaluesReducedPseudopotetialMesh} \end{figure*}

As a final point, we also investigate the effect of truncating the Fourier coefficients of the \textit{in-situ} pseudopotential, to gauge its smoothness. In Fig.~\ref{Fig:EigenvaluesReducedPseudopotetialMesh} we show results of energy bands of bcc Na, when the \textit{in-situ} pseudopotential, being calculated from Eq.10, has its higher Fourier components truncated. In practice this means keeping from the original \textit{in-situ} pseudopotential, only components from a $N\times N\times N$ mesh (that is smaller than the origial mesh of $11\times 11\times 11$). It is interesting to note from Fig.~\ref{Fig:EigenvaluesReducedPseudopotetialMesh} that good agreement between full-potential, all-electron theory and \textit{in-situ} pseudopotential theory can be is achieved for all considered Fourier meshes, except the very smallest one ($3\times 3\times 3$). This indicates that equal accuracy to all-electron theory can be achieved from an \textit{in-situ} pseudopotential theory, that is represented by only 125 ($5\times 5\times 5$) plane waves.

\section{Discussion and Conclusion} \label{Sec:Discussion}

In this paper, we have demonstrated a proof-of-concept of how it is possible to
calculate a pseudopotential from an all-electron, electronic structure method. 
For reference, in Fig.~\ref{Fig:Eigenvalues}~c,~d we also consider for a much cruder choice of core function, using $f(r) = 1$, $R_0 = 0.55R_{MT}$ and $R_1 = 0.75R_{MT}$. The agreement of the eigenvalues is in fact surprisingly good also in this case, even though the pseudowavefunction differs significantly from that of the Quantum Espresso in the core region.
Technically, this amounts to solving the inverse Kohn-Sham equation for one or
a few eigenvalues and eigenstates, which have been obtained from the
all-electron theory. The method proposed here relies on replacing the rapidly
oscillating part of an eigenstate close to the nucleus (in the muffin-tin
sphere) with a smoother and much softer form, which allows for fast convergence
in the expansion of the Fourier series. In principle, this method is not
restricted to using the solid state as the reference electronic configuration
and can be readily extended to molecular species or crystal surfaces. It also
does not require the use of LMTOs as basis functions and is, for example, also
suitable for the LAPW or LCAO basis set. It is also not mandatory to use the zone center to evaluate the \textit{in-situ} pseudopotential, other points of the Brillouin zone can be used, and it is possible to take averages of \textit{in-situ} pseudopotentials from several points, to get a final \textit{in-situ} pseudopotential to use for further studies.

In this work, we limit our discussion to the construction of the local
components of the pseudopotentials. Its extension to the nonlocal components~\cite{PhysRevB.26.4199, PhysRevLett.48.1425, PhysRevB.41.7892}
is natural, and necessary in order to resolve higher lying energy bands, an effort which represents ongoing work. Even without the nonlocal components,
the proposed \textit{in-situ} pseudopotential method reproduces energy dispersion of the
3$s$-like band states to good accuracy throughout the Brillouin zone. The
largest discrepancy between the \textit{in-situ} pseudopotential outlined here, and results from an
all-electron method, is at the Brillouin zone boundary. This is expected since
these zone-boundary states have an admixture of angular momentum characters
that can only be properly described with the inclusion of the nonlocal
components in the \textit{in-situ} pseudopotential. Furthermore, the crystal momentum of these states is the farthest away from the k-point at which the \textit{in-situ} pseudopotential was calculated.

It is well-known that the transferability of an atomic pseudopotential
can be systematically improved, by reducing $r_c$ at the cost of greater
computational cost~\cite{PhysRevB.26.4199}. With the construction of the
pseudopotential \textit{in situ}, using the native state as the reference, this
requirement of transferability can potentially be relaxed if an \textit{in-situ} pseudopotential is used. This may even allow for a
larger pseudopotential radii cutoff for the same
convergence, thereby reducing the number of Fourier components needed in the
series expansion and a reduced computational cost. 
This method also automatically takes into account the nonlinear~\cite{Louie1982} nature of the exchange and 
correlation interaction between the core and valence charge densities,
which is important when a valence-only pseudopotential is used for an akali metal 
like Na~\cite{Louie1982}, which only has one
electron in the valence shell. In a typical calculation using
atomic pseudopotentials, these interactions are first assumed to be linear, before
adding the nonlinear core contributions as a perturbative correction. Another benefit of generating the
pseudopotential in the native solid-state environment is that basis-set
convergence is already controlled at the level of the all-electron calculation.
For example, if the LMTO-basis set is used for the all-electron calculation (as
in our case), convergence parameters will include the number of
Fourier-components to match LMTOs, as well as core-leakage that will indicate
if certain semi-core states have to be treated as valence states. Computational
cost versus accuracy can then be optimized.

The methodology suggested here can also be extended to include the spin-polarized
case. One must then keep track of spin-indices of the all-electron generated
eigenvalues and eigenvectors in the analysis presented in
Sec.~\ref{Sec:Method}. Following the steps in the methods section that
describe the pseudopotential generation, one could then obtain \textit{in-situ} pseudopotentials
for spin-up states and spin-down states separately. After unscreening of these
pseudopotentials (removing contributions from exchange and correlation of the
electron gas, as well as the Hartree potential) one would obtain spin-dependent
pseudopotentials that are able to accurately reproduce magnetic moments
and spin-dependent information of all-electron theory.  Spin-orbit effects may
also be incorporated in the proposed \textit{in-situ} pseudopotentials, since
the method outlined can be used for any spin-orbit calculated eigenstate. Finally, we speculate that the \textit{in-situ}
pseudopotential can be generalized, such that effects from a self-energy,
$\Sigma^{\mathbf{k}}$, (e.g., obtained from the $GW$ approximation or the
dynamical mean field theory) are incorporated in the pseudopotential. This could be achieved by
associating $\tilde{\epsilon}_{\mathbf{k}n}$ with
$\tilde{\epsilon}_{\mathbf{k}n}+Re\Sigma_{\mathbf{k}}$ in Eq.~\ref{Eq:SE}. The
steps outlined above represent an investigation that is underway.

\section{Acknowledgement} O.E. acknowledges support from the Swedish Research
Council, the Knut and Alice Wallenberg foundation, the Foundation for Strategic
Research, the Swedish Energy Agency, the European Research Council (854843-FASTCORR), and eSSENCE. Calculations made on the
SNIC infrastructure for high performance computing. 

\newpage

\bibliography{main}
\bibliographystyle{naturemag}

\end{document}